\documentclass[aps,prl,twocolumn,footinbib,superscriptaddress,amssymb,letterpaper]{revtex4}
\usepackage{color}
\usepackage{epsfig}
\usepackage{graphicx}
\usepackage{epstopdf}
\usepackage{mathtools}
\usepackage{bbold}
\usepackage{feynmp}
\usepackage{slashed}
\usepackage{subfigure}
\usepackage{amsmath}
\usepackage{amsfonts}
\usepackage{amssymb}
\usepackage{color}
\usepackage{mathrsfs}
\usepackage[colorlinks=true, linkcolor=blue, citecolor=blue, urlcolor=blue]{hyperref}
\usepackage{tikz}
\usepackage{pgfplots}
\def\be{\begin{equation}}
\def\ee{\end{equation}}

\begin{document}

\title{Entanglement Dynamics in 2D CFT with Boundary:\\[2mm] {Entropic origin of JT gravity and Schwarzian QM}}

\def\spc{\hspace{.5pt}}

\author{Nele Callebaut} 
\email{nelec@princeton.edu}
\affiliation{Department of Physics, Princeton University, Princeton, NJ 08544, USA}
\affiliation{Department of Physics and Astronomy, Ghent University, Krijgslaan 281-S9, 9000 Gent, Belgium}

\author{Herman Verlinde}
\email{verlinde@princeton.edu}
\affiliation{Department of Physics, Princeton University, Princeton, NJ 08544, USA}

\date{\today}

\begin{abstract}

We study the dynamics of the geometric entanglement entropy 
of a 2D CFT in the presence of a boundary. We show that this dynamics is governed by local equations of motion, that take the same form as 2D Jackiw-Teitelboim gravity coupled to the CFT. If we assume that the boundary has a small  thickness $\epsilon$ and constant boundary entropy, we derive that its location satisfies the equations of motion of Schwarzian quantum mechanics with coupling constant $C = {c\spc\epsilon}/{12\pi}$. We rederive this result via energy-momentum conservation.
\end{abstract}

\def\calO{{b}}
\def\be{\begin{equation}}
\def\ee{\end{equation}}
\def\Cz{\delta C}
\def\Cb{C}


\maketitle
\def\mathbi#1{\textbf{\em #1}} 
\def\som{{ \textit{\textbf s}}} 
\def\tom{{ \textit{\textbf t}}} 
\def\nom{n} 
\def\mom{m}
\def\la{\langle}
\def\bea{\begin{eqnarray}}
\def\eea{\end{eqnarray}}
\def\is{& \! = \! & }
\def\ra{\rangle}
\def\half{{\textstyle{\frac 12}}}
\def\cL{{\cal L}}
\def\halfi{{\textstyle{\frac i 2}}}
\def\ba{\begin{eqnarray}}
\def\ea{\end{eqnarray}}
\newcommand{\rep}[1]{\mathbf{#1}}
\newcommand{\Tr}{\, {\rm Tr}}
\def\uU{\bf U}
\def\be{\bea}
\def\ee{\eea}
\def\delbar{\overline{\partial}}
\newcommand{\smpc}{\hspace{.5pt}}
\def\ra{\bigr\rangle}
\def\la{\bigl\langle}
\def\ccdot{\!\spc\cdot\!\spc}
\def\nspc{\!\spc\smpc}
\def\tr{{\rm tr}}

\addtolength{\baselineskip}{0mm}
\addtolength{\parskip}{.1mm}
\def\ra{\bigr\rangle}
\def\la{\bigl\langle}
\def\li{\bigl|\spc}
\def\ri{\bigr |\spc}

\def\hf{\textstyle \frac 1 2}

\newcommand{\beq}{\begin{equation}}
\newcommand{\eeq}{\end{equation}}
\newcommand{\nn}{\nonumber\\} 
\newcommand{\lb}{{\langle}}
\newcommand{\rb}{{\rangle}}
\def\spc{\hspace{1pt}}
\def\is{& \! = \! &}
\def\d{{\partial}}
\def\n{{\bf \widehat n}}
\def\k{{\bf k}}
\newcommand{\na}{\mbox{\boldmath$\nabla$}}
\newcommand{\om}{\omega}
\newcommand{\calo}{{\cal O}}
\newcommand{\calL}{{\cal L}}
\newcommand{\vphi}{{\varphi}}
\newcommand{\zb}{\overline{z}}
\def\sS{{S}}
\setcounter{tocdepth}{2}
\addtolength{\baselineskip}{0mm}
\addtolength{\parskip}{0mm}
\addtolength{\abovedisplayskip}{0mm}
\addtolength{\belowdisplayskip}{0mm}
\def\ssigma{\mbox{\small $\tilde{S}_0$}}
\def\cS{{\cal S}}
\def\xx{{\smpc \rm x}}
\subsection{Introduction} 
\vspace{-3mm}

\def\X{\xx}

It has been known for some time that there is a deep connection between the equations of motion of gravity 
and the dynamical properties of entanglement \cite{Jacobson:1995ab}. This connection was recently used to identify the quantum null energy condition (QNEC) as a pure QFT result, by taking a $G_N\to 0$ decoupling limit of a corresponding dynamical property of semi-classical general relativity 
\cite{Bousso:2015mna,Balakrishnan:2017bjg}. 
The QNEC gives a bound on the change in entanglement entropy in terms of the  energy-momentum flux. 
However, there are indications that, in rather general settings, this inequality may in fact be saturated and reduce to the first law of entanglement thermodynamics  \cite{Leichenauer:2018obf,Koeller:2017njr}. 

In this note, we study this connection in the context of 2D CFT with a boundary.  We show that the first law of entanglement thermodynamics in this set-up takes the same form as the equations of motion of 2D Jackiw-Teitelboim (JT) gravity \cite{Jackiw:1984je,Teitelboim:1983ux}, {and discuss the relation with the 2D QNEC.}  
The equations that establish the entropic interpretation of JT gravity are not new. However, by viewing them from a bottom-up perspective (we don't take gravity as input but aim to obtain it as output) gives rise to a new derivation of JT gravity and of Schwarzian quantum mechanics \cite{KitaevTalks,Maldacena:2016hyu,Jensen:2016pah,Maldacena:2016upp,Engelsoy:2016xyb} from purely entropic considerations.

\begin{figure}[h!]
\begin{center}
\begin{tikzpicture}[scale=1.05]
\draw[color={green}][thick] (0,0) -- (2,0);
\draw[color={red}][thick] (0,0) -- (-2.5,0);
\draw[thick] (2,-2.3) -- (2,2.3);
\draw[dashed] (-2,2) -- (2,-2);
\draw[dashed] (-2,-2) -- (2,2);
\draw[fill=black] (0,0) circle (0.08);
\draw (-0,-.5) node {\footnotesize $(u,v)$};
\draw[color={blue!20}] plot[variable=\t,samples=1000,domain=-33:33] ({2.5* sec(\t)-1.08},{2.5* tan(\t)});
\draw[color={blue!20}] plot[variable=\t,samples=1000,domain=-56:56] ({1.14* sec(\t) -.23},{1.1* tan(\t)});
\draw[color={blue!20}] plot[variable=\t,samples=1000,domain=-78:78] ({.35* sec(\t)+.04},{.35* tan(\t)});
\draw[color={blue}] plot[variable=\t,samples=1000,domain=-33:33] ({-2.5* sec(\t)+1.08},{-2.5* tan(\t)});
\draw[color={blue}] plot[variable=\t,samples=1000,domain=-56:56] ({-1.14* sec(\t) +.23},{-1.1* tan(\t)});
\draw[color={blue}] plot[variable=\t,samples=1000,domain=-78:78] ({-.35* sec(\t)-.04},{-.35* tan(\t)});
\end{tikzpicture}
\end{center}
\vspace{-2mm}
\caption{The interval labeled by $(u,v)$ (horizontal green line) and its complement (horizontal red line) and the corresponding modular flow (indicated in blue). }
\vspace{-3mm}
\end{figure}
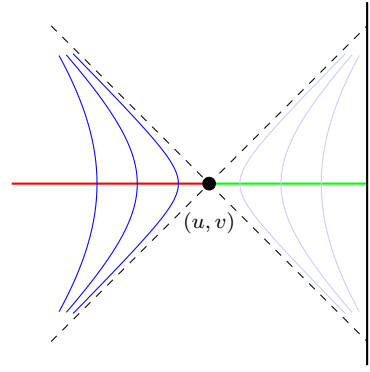

Consider a 2D CFT with central charge $c\!\gg\! 1$ defined on the $u,v$ plane in the presence of a boundary located~at
\ba
\label{boundary}
{\rm Boundary}\spc \is\spc \{ u = v \equiv t\} . 
\ea
Due to the presence of the boundary, we can associate to a given point $(u,v)$ a space-like interval between the point and the boundary, as indicated by the green line, and its space-like complement, indicated by the red line.

Let $\rho_{(u,v)}$ 
denote the density matrix  of the CFT on the red half line, 
obtained by tracing out the green segment. From $\rho_{(u,v)}$, we extract the entanglement entropy $\sS(u,v)$
and modular Hamiltonian $K(u,v)$ via
\ba
\sS(u,v) \is -\Tr\bigl(\rho_{(u,v)} \log \rho_{(u,v)}\bigr) \spc 
\\[1mm]
\rho{(u,v)} \is \frac 1 {Z(u,v)} {e^{-K(u,v)}}
\ea
with $Z(u,v) \! =\! \Tr(e^{-K(u,v)})$.
Both quantities are local,  in the sense that they are 
functions of a single space-time point $(u,v)$. They are related via
\ba
\sS(u,v) \is \lb K(u,v) \rb - F(u,v), \nonumber\\[-2mm]\label{skf}\\[-2mm]\nonumber
F(u,v) &\! \equiv \! &  - \log  Z(u,v).
\ea
Since $\delta F = \lb \delta K \rb$ under infinitesimal variations of the state, we deduce the first law of entanglement thermodynamics 
\ba
\label{firstlaw}
\delta S(u,v) \is \delta \bigl\langle K(u,v) \bigr\rangle,
\ea
where  the variation acts only on the expectation value.
\vspace{-4mm}

\subsection{JT gravity from entanglement}
\vspace{-4mm}

We now show that, for a general class of states defined below, the entropy $S(u,v)$ and modular Hamiltonian $K(u,v)$ satisfy the same local equations of motion as Jackiw-Teitelboim gravity.  

We introduce a second coordinate system $(\xx^+,\xx^-)$  related to $(u,v)$ via a general 2D conformal transformation 
\ba
(u,v) \, \to \, (\xx^+(u) \!\!&\!\!\smpc,\!&\!\! \xx^-(v))
\ea
that leaves the location of the boundary  \eqref{boundary} fixed at $\xx^+ = \xx^-$. Hence the conformal transformation is parametrized by a single reparametrization mode along the boundary
\ba
\label{bv}
\xx^+(t) \is \xx^-(t) \, = \, \tau(t)  .
\ea
Now consider the class of states of the form
\ba
|\spc \Psi \spc \rb \spc \is \mbox{a few operators acting on}\; |\spc 0 \spc \rb_{\! \xx}\nonumber\\[2mm]
|\spc 0 \spc \rb_{\! \xx} \is \mbox{vacuum in $\xx^\pm$ coordinate system}.\nonumber
\ea
We assume that the operators create only low energy excitations. The vacuum state $|\spc 0 \spc \rb_{\! \xx}$ depends on the distinction between positive and negative frequencies, and thus depends on the choice of coordinates. Expectation values with respect to~$| 0 \rb_{\! \xx}$ are equal to the CFT path integral with corresponding asymptotic vacuum boundary conditions.

The entanglement entropy $S(u,v)$ associated to the state $|\Psi\rb$ on the red half-line in Figure 1 bounded by the point $(u,v)$ can be decomposed into a vacuum contribution and a first order correction
\ba
S(u,v) \is S_0(u,v) + S_1(u,v).
\ea
The vacuum term  is the leading contribution for large central charge $c$, and takes the form \cite{Holzhey:1994we,Cardy:2016fqc}
\ba
\label{szero}
S_0(u,v) \is  \frac{c}{12} \log\left(\frac{(\xx^+\!\!\smpc -\!\smpc\xx^-)^2}{4 \epsilon^2 \partial_u \xx^+ \partial_v \xx^-} \right)
\ea
where $\epsilon$ is the UV cutoff in $(u,v)$ coordinates with standard Minkowski metric $ds^2 = -du dv$. Here we take it as fact that the vacuum entanglement entropy of the red half line and green segment in Figure 1 are both equal \footnote{Here we omit a possible constant term given by the boundary entropy. The boundary entropy and its dynamics will be discussed in the next subsection.}.

A first link with JT gravity relies on the identification of the rescaled entropy 
\be
\label{sscale}
{{\tilde S}_0(u,v)} \is  \frac{12}{c}\,S_0(u,v) - \log\left(\frac {\ell^2} {\epsilon^2}\right)
\ea
with the conformal mode of a constant curvature metric
\ba
\label{ccmetric}
ds^2 \is - e^{-{\tilde S}_0(u,v)} du dv \, = \, - \frac{4 \ell^2 d\xx^+d\xx^-}{(\xx^+-\xx^-)^2} 
\ea
with uniformizing coordinates $(\xx^+,\xx^-)$. In \eqref{sscale}
we also applied a constant shift to replace the UV cut-off by some finite length scale $\ell$, so that  the rescaled vacuum entanglement $\ssigma$ satisfies the Liouville equation of motion~\cite{deBoer:2016pqk}
\ba
\label{liouv}
\qquad \partial_u \partial_v \ssigma(u,v) \is \frac \Lambda 4 \spc e^{-{\tilde S}_0(u,v)}, \qquad \Lambda \!\smpc =\! \smpc \frac{2}{\ell^2}
\ea
with finite cosmological constant.  The  vacuum state $| 0 \rb_{\! \xx}$ and the  metric \eqref{ccmetric} share the same global isometries. 

We identify the vacuum contribution $S_0$ to the entropy with the free energy term $F(u,v)$ in equation \eqref{skf}. We thus define the modular Hamiltonian such that its vacuum expectation value vanishes.
The subleading term 
\ba
\label{firstlaw}
\sS_1(u,v) \is \lb K(u,v) \rb
\ea 
is the contribution to the entanglement entropy due to the variation away from the vacuum state. Here $\lb ... \rb$ denotes the expectation value with respect to $|\Psi\ra$. We will now show that $\sS_1(u,v)$ satisfies the same equation of motion as the dilaton field in JT gravity.

In a 2D CFT, the modular Hamiltonian of a vacuum state $|0\rb_\xx$ is given by the generator of time evolution along the Killing vector 
of the constant curvature metric \eqref{ccmetric} that leaves the point $(u,v)$ fixed. This modular flow is indicated by the blue lines in Figure 1. The modular Hamiltonian takes the form \cite{Casini:2011kv}
\ba
\label{modham}
& &\!\!\!\!\!\! K(\xx)\spc = \spc  
K_+(\xx) \spc + K_-(\xx)\nonumber \\[-2.2mm]\\[-2mm] \nonumber
K_\pm(\xx) \is \pm 2\pi \int_{\pm\infty}^{\xx^\pm} \!\!\!\!\! ds \, \frac{(\xx^+\!\!-\!\smpc s)(\smpc s-\!\xx^-)}{\xx^+-\xx^-} \,T_{\pm\pm}(s)
. 
\ea
Here $T_{\pm\pm}$ denotes the CFT energy-momentum tensor in $(\xx^+\!,\xx^-)$ coordinates. It is related to the energy-momentum tensor in $(u,v)$ coordinates via
\ba
\label{uustress}
T_{uu} = \spc -\frac{c}{24\pi} \, {\rm Sch}(\xx^+,u) \, + \, T_{++} \Bigl(\frac{d \smpc \xx^+\!}{du}\spc\Bigr)^2, \qquad  & & \\[2mm]
{\rm Sch}(\xx^+\!,u)  \spc = \spc  \frac{\dddot{\xx}^+}{\ddot{\xx}^+} \!-\frac 3 2  \Bigl(\frac{\ddot{\xx}^+\!}{\dot\xx^+\!}\Bigr)^2 =  -\frac 1 2( \partial_u \ssigma)^2\! - \! \partial_u^2 \ssigma  
& &  \label{schw}
\ea
with $\dot{\xx}^+ = \partial_u \xx^+$, etc. The first term on the r.h.s. in \eqref{uustress} denotes the vacuum contribution, which for large $c$ dominates over the second term. 
The modular Hamiltonian depends only on the second, subleading term. 

The modular Hamiltonian \eqref{modham} satisfies the second order differential equation
\ba 
\label{Keq}
\nabla_{\! \smpc +} \partial_+ K(\xx) \is 2\pi\smpc T_{++}(\xx^+)\nonumber \\[-1.75mm]\\[-1.75mm]
\nabla_{\! \smpc -} \partial_- K(\xx) \is 2\pi\smpc T_{--}(\xx^-)\nonumber 
.
\ea
Here the covariant derivative is taken with respect to the constant curvature metric \eqref{ccmetric}.
Via the first law of entanglement thermodynamics \eqref{firstlaw}, we thus deduce that $S_1(\xx)$ also satisfies a second order differential equation
\ba
\nabla_{\! \smpc +} \partial_+  S_1(\xx) \is 2\pi\smpc \bigl\langle T_{++}(\xx) \bigr\rangle\nonumber \\[-1.75mm]\label{qneclim} \\[-1.75mm]
\nabla_{\! \smpc -} \partial_-   S_1(\xx) \is 2\pi\smpc \bigl\langle T_{--}(\xx) \bigr\rangle\nonumber
.
\ea
The above two equations for $K(\xx)$ and $S_1(\xx)$ are identical to the equations of motion of the dilaton in JT gravity. So by identifying $K(\xx)$ with the quantum dilaton field operator, and $S_1(\xx)$ with its semi-classical expectation value, we have now derived JT gravity from pure entanglement dynamics in 2D CFT. The diagonal JT equations  
\ba 
\partial_+ \partial_- K(\xx) +\frac\Lambda 4 \spc e^{-\tilde{S}_0} K(\xx)\is 0\nonumber\\[-2mm]
\label{diagJT}\\[-2mm]
\partial_+ \partial_-  S_1(\xx) +\frac \Lambda 4 \spc e^{-\tilde{S}_0} S_1(\xx)\is 0 \nonumber
\ea
are also satisfied. 
For a Rindler modular Hamiltonian, the local relation between its second derivative  and the stress tensor was already observed in \cite{Wall:2011hj} and \cite{Leichenauer:2018obf}.  

\smallskip

Some further comments are in order. 
\smallskip

1. Equations \eqref{qneclim}, \eqref{uustress}-\eqref{schw} together with \eqref{szero}  combine into the first order variation, in the perturbation away from the vacuum, of the following non-linear equation for the total entropy $S =S_0 + S_1$
\ba
\label{uvqnec}
\partial_u^2 S\!\smpc +\!\smpc \frac{\raisebox{-.5pt}{$6$}} {\raisebox{.5pt}{$c$}}\spc (\partial_u S)^2 \!  \is\!\smpc 2 \pi\smpc \bigl\langle T_{uu}\bigr\rangle ,
\ea
or equivalently, the following non-linear equation for $S_1$
\ba
\label{dqnec}
\nabla_{\! \pm}\partial_\pm S_1\! +\!\smpc  \frac{\raisebox{-.5pt}{$6$}}{\raisebox{.5pt}{$c$}}\spc  (\partial_\pm S_1)^2\!\!\smpc \is\!\smpc 2 \pi\smpc \bigl\langle T_{\pm\pm}\bigr\rangle .
\ea
Notice that \eqref{uvqnec} looks like the Liouville energy-momentum tensor in flat (u,v) coordinates,  
while the linearized equation \eqref{dqnec} looks like a covariant equation in the constant curvature metric \eqref{ccmetric}. Hence the effective background metric \eqref{ccmetric} arises due to the linearization around a given vacuum entropy $S_0$.

\smallskip

2. Equation \eqref{dqnec} and its linearized approximation~\eqref{qneclim} coincide with the saturation limit of the 2-dimensional QNEC \cite{Bousso:2015mna,Wall:2011kb,Wall:2018ydq,Khandker:2018xls}. 
The QNEC depends on a choice of affine parameter along a null surface, which in turn is linked to a choice of (local) vacuum state and a choice of metric.  For the CFT vacuum state at hand, this metric is given by the constant curvature metric \eqref{ccmetric}. The form of the metric is fixed by the requirement that it must be symmetric under the M\"obius group 
$\xx^\pm \to \frac{ a\xx^\pm + b }{c\xx^\pm + d}.$

\smallskip 

3. The linearized equation \eqref{qneclim} has a more general solution \cite{Almheiri:2014cka}, which includes an extra zero mode
\ba
\label{entone}
S_1(\xx) \! \is \!  
\spc\langle K_+(\xx)\rangle +  \langle K_-(\xx) \rangle\spc - \spc  \frac{4\pi \spc \delta C}{\!\xx^+\! -\xx^-\!\!} 
\ea
proportional to a constant $\delta C$ with dimension of length.  This extra term looks unfamiliar from a 2D CFT perspective, as it appears to break M\"obius invariance. 
However, we can interpret it as the result of a small misalignment between the location of the boundary and the $\xx$-coordinates that fix the vacuum state $|0\rangle_{\! \xx}$: performing an infinitesimal coordinate shift  
\ba
\label{xshift}
(\xx^+\! ,\xx^-) \to (\xx^+\!\!\spc - \delta \epsilon, \xx^-\!\!\spc + \delta \epsilon) 
\ea 
in the expression \eqref{szero} for the vacuum entanglement entropy $S_0$ produces an extra term 
of the same form as the zero mode in \eqref{entone} with 
\bea 
\label{dc}
\delta C \is \frac{c }{12\pi}\spc \delta\epsilon.
\eea
The negative sign in \eqref{entone} means that we are removing a small amount of entropy by shifting the boundary inward. 
In the next subsection, we will make use of this observation to derive the position dependence of the boundary effective action.

\medskip

Let us summarize. We have derived the JT equations  from pure CFT considerations. In particular, we did not use or establish an action principle based on a dilaton gravity action. We could decide, however, to summarize the entanglement dynamics of 2D CFT in terms of an effective action. This action would then take the JT form 
\ba
\label{effact}
\cS_{\rm eff} \is \frac{1}{16\pi} \int\! d^2 x\, \Phi(R + \Lambda) \, + \, \cS_{\rm CFT}
\ea
via the identification between the leading and subleading contribution $\sS_0$ and $\sS_1$ to the entanglement entropy with respectively the scale factor 
\eqref{sscale} of the metric \eqref{ccmetric} with curvature $R$, 
and the dilaton  
\ba 
\label{dilK}
\mbox{\large $\frac 1 4$} \spc \Phi(u,v) \is -S_1(u,v) = -\lb K(u,v)\rb . 
\ea 
Note that JT gravity has no physical degrees of freedom: its equations of motion \eqref{qneclim} are imposed as gauge constraints that define $S_1$ as a collective mode of the CFT. 

\vspace{-2mm}

\subsection{Entropic derivation of Schwarzian QM}

\vspace{-3mm}

The boundary \eqref{boundary} has a fixed location in $(u,v)$ coordinates. In $\xx$-coordinates, its location is specified by a single function $\tau(t)$ of time \eqref{bv}. 
A priori, one would think that we should be free to choose the $\xx$-coordinates and $\tau(t)$ in any way we want. However, the boundary introduces a non-trivial coupling between the left- and right-moving sector of the CFT. So we can not arbitrarily choose the local CFT vacuum state $|0\rb_\xx$, since the left- and right-moving vacuum conditions are correlated via reflection at the boundary. We want to derive equations that specify this correlation.

We will show that the effective 1D theory for the boundary trajectory is given by Schwarzian QM. We will give two independent derivations: one based on entropy and one based on energy-momentum conservation. Both derivations require one additional physical input: we will assume that the boundary has a small but finite thickness $\epsilon$, as indicated in Figure 2 by the dashed blue line. The boundary is a straight line $(u,v) = (t+\epsilon,t-\epsilon)$ in the original ($u,v$) coordinates.    One can think of $\epsilon$ as a UV regulator.
The $(\xx^+,\xx^-)$ coordinates at the cut-off location satisfy 
\bea
\label{bbv}
\xx^\pm(t \pm \epsilon) \is \tau(t) \pm \spc\epsilon \spc  \dot{\tau}(t) \, .
\eea
The vacuum entanglement entropy $S_0$ in \eqref{szero}  identically vanishes along the boundary 
\ba
\label{szeroc}
S_0(t+\epsilon, t-\epsilon) \is 0 
\ee 
up to terms of order $\epsilon^2$. We can view this condition as the definition of the boundary location. Note that this condition introduces a length scale, and thus explicitly breaks 2D conformal invariance. We should thus expect that the boundary reparametrization mode $\tau(t)$ will become a dynamical `pseudo-goldstone mode'. We want to derive its effective action.

We will think of the cut-off length scale $\epsilon$ as a Wilsonian RG scale, introduced by integrating out the UV degrees of freedom very close to the boundary. Following this intuition, we will derive the explicit form of the boundary effective action by considering the effect of a small variation $\epsilon \to \epsilon + \delta\epsilon$ in the cut-off scale.   This variation amounts to moving the location of the dynamical boundary  in equation \eqref{bbv} by a small amount $2\delta \epsilon  \spc \dot{\tau}(t)$. From this, one directly verifies that the cut-off variation preserves the condition \eqref{szeroc}. Hence it looks like the vacuum entanglement entropy transforms trivially under the RG flow. However, thus far we have only done the trivial part of the full RG step, i.e. change the size of our measuring stick.

The non-trivial part of the RG step involves integrating out the CFT degrees of freedom inside the small extra boundary layer~$\delta \epsilon$. 
Here by `integrating out' we do not mean `tracing over' but `remove and project all degrees of freedom that were entangled with the small boundary layer onto a pure state'. This step is reminiscent of cMERA \footnote{
	The entanglement renormalization approach cMERA \cite{Haegeman:2011uy} provides a variational, real-space implementation of Wilson's momentum-shell RG picture, with the difference that coarse-graining is achieved by projecting degrees of freedom out, rather than tracing   
	them out. In practice this is implemented by a scaling action   
	that maps the to be discarded degrees of freedom 
	outside the physical range. 
	The coarse-graining in cMERA is preceded by a disentangling action that removes short-range entanglement. We imagine the same happens here.   
	}. 
This way, we ensure that the whole system remains in a pure state. We propose that this RG step can be accounted for by applying the active 
coordinate shift \eqref{xshift} in the expression \eqref{szero} for the vacuum contribution to the entanglement entropy. 
This adds the extra zero mode term \eqref{entone} to the $S_1$ contribution of the entropy, with $\delta C$ given in \eqref{dc}. As mentioned above, this contribution is negative because, by projecting out the boundary layer, we are removing entanglement. The total amount of removed entanglement entropy $\delta \sS_b(t)$ is equal to the value of $S_1(u,v)$ at the cut-off location 
\ba
\delta \sS_b(t) \! &\equiv& \! \sS_1( t + \epsilon, t-\epsilon).
\ea
We now adopt the adiabatic postulate that the total removed entropy is constant in time. In other words, we assume that the entropy in- and out-flux at the boundary is negligible compared to the energy flux. We will see that the natural value for this constant boundary entropy~is
\ba
\label{boxedo}
\delta \sS_b \is -\frac{c}{6} \spc \frac{\delta \epsilon}{\epsilon}\,=\, - \frac{4\pi \spc \delta C}{2\epsilon} .
\ea
The sign of the boundary entropy indicates that it accounts for the removal of boundary degrees of freedom within $\delta \epsilon$.

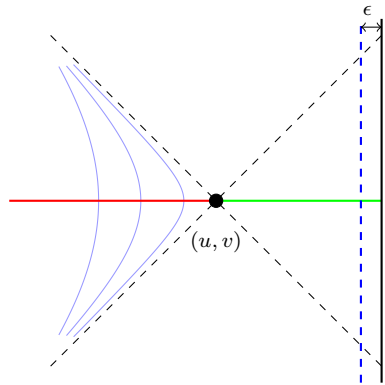
\begin{figure}[t!]
\begin{center}
\begin{tikzpicture}[scale=1.1]
\draw[color={blue!40}] plot[variable=\t,samples=1000,domain=-33:33] ({-2.5* sec(\t)+1.08},{-2.5* tan(\t)});
\draw[color={blue!40}] plot[variable=\t,samples=1000,domain=-56:56] ({-1.14* sec(\t) +.23},{-1.1* tan(\t)});
\draw[color={blue!40}] plot[variable=\t,samples=1000,domain=-78:78] ({-.35* sec(\t)-.04},{-.35* tan(\t)});
\draw[color={green}][thick] (0,0) -- (2,0);
\draw[color={red}][thick] (0,0) -- (-2.5,0);
\draw[thick][color={blue}][dashed] (1.75,-2.2) -- (1.75,2.2);
\draw[thick] (2,-2.2) -- (2,2.2);
\draw[<->] (2,2.1) -- (1.75,2.1);
\draw (1.83,2.3) node {\footnotesize $\epsilon$};
\draw[dashed] (-2,2) -- (2,-2);
\draw[dashed] (-2,-2) -- (2,2);
\draw[fill=black] (0,0) circle (0.08);
\draw (0,-.5) node {\footnotesize $(u,v)$};
\end{tikzpicture}
\end{center}
\vspace{-2mm}
\caption{We assume that the boundary layer has a finite effective thickness $\epsilon$, as indicated by the blue   
dashed line. 
}
\vspace{-2mm}
\end{figure}

The derivation of the Schwarzian action is now identical to \cite{Engelsoy:2016xyb}.  We summarize the main steps.
Inserting the parametrization \eqref{bbv} into the explicit form \eqref{entone} of the entanglement entropy $\sS_1$ gives 
\ba
\delta \sS_b(t)  \is \smpc \frac{-4\pi \Cz -  I_+(t) +  I_-(t) }{2 \epsilon\spc \dot\tau(t)}\, = \, - \frac{4\pi \spc \delta C}{2\epsilon}
 \label{eq:e105}\\[2mm]
I_\pm(t)  \is\spc 2\pi \spc \int_{\pm\infty}^{\tau(t)}\!\!\!\! d s\; (s - \tau(t))^{2}\,  \lb T_{\pm\pm}(s) \rb. 
\ea
The first of these equations implies that the boundary trajectory satisfies the first order differential equation
\bea
\label{zdynam}
4\pi \Cz \spc {\dot{\tau}(t)}\! \is\! \spc 4\pi \Cz\spc + \spc   I_+(t)  \spc -  I_-(t)  \spc . \
\eea
In the vacuum state $I_\pm(t) = 0$ and this equation reduces to $\dot{\tau}(t) = 1$. The condition that $\tau =t$ in the vacuum
fixes the identification~\eqref{boxedo}.
 
Using that $ \mbox{${d^3 I_\pm}/{d\tau^3}$} \, = \, 4\pi  \lb T_{\pm\pm}\rb  $, we can write \eqref{zdynam} in local form as a fourth order differential equation
\ba
\label{jerk}
\Cz \, \frac{d^{ 3} \dot{\tau}}{d\tau^{\smpc 3}}\,  \is  \lb T_{++} \rb_\epsilon \spc - \, \lb T_{--} \rb_\epsilon.
\ea
Here we added the subscripts to indicate that the expectation values on the right-hand side are defined in the theory with boundary cut-off $\epsilon$.
Equation \eqref{jerk} can be recognized \cite{Engelsoy:2016xyb} as the equation of motion of a 2D CFT coupled to the Schwarzian  action
\ba
\label{rgs}
{\cal S}^{\epsilon +\delta \epsilon}_{\rm CFT} \,=\, \cS^{\spc \epsilon}_{\rm CFT} \! & - &\!  \Cz \int\! dt \, {\rm Sch}(\tau,t) \, .
\ea
with coupling $\delta C$. Here $\cS^{\epsilon}_{\rm CFT}$ denotes the CFT effective action with boundary cut-off $\epsilon$, defined such that its stress tensor is given by $\lb T_{\pm\pm}\rb_\epsilon$.  
We view the  identity \eqref{rgs} as an RG equation for the CFT effective action $S^{\spc \epsilon}_{\rm CFT}$ with cut-off~$\epsilon$. Integrating with respect to $\epsilon$, we deduce that the full effective action of the boundary trajectory is given by the Schwarzian action 
\ba
\label{schwqm}
{\cal S}^{\spc \epsilon}_{\rm CFT} \is {\cal S}^{\epsilon \to 0}_{\rm CFT}\, - \, C \int\! dt \, {\rm Sch}(\tau,t), 
\ea
with coupling constant 
\ba
C\spc  \is \spc \frac{c\spc \epsilon}{12 \pi} \, .
\ea

At the level of equations, the above derivation of the Schwarzian action is not new.
The new observation is in the arrow of implication -- the viewpoint that Schwarzian dynamics can be derived from purely entropic considerations. We will further comment on the meaning of equation \eqref{schwqm} in the concluding section.

\vspace{-2mm}

\subsection*{Energy-momentum conservation}
\vspace{-3mm}

We will now rederive the boundary effective action \eqref{schwqm} from energy-momentum conservation.
It will be convenient to define the boundary coordinate $\tau(t)$ as the proper length along the boundary 
\ba 
\label{proper}
{d \X \over d \tau}^{\! +}{d \X \over d \tau}^{\! -} \! \is 1.
\ea
For small $\epsilon$, this definition of $\tau(t)$ is  consistent with equation \eqref{bbv}. We assume that the CFT boundary state 
satisfies
perfectly reflecting Ishibashi boundary conditions in $(u,v)$ coordinates, so that $\lb T_{uu} \rb= \lb T_{vv}\rb $. In terms of $\xx$-coordinates, this reflection condition contains an extra contribution from the conformal anomaly
\ba
\label{reflect}
\lb T_{--} \rb  
\is \lb T_{++} \rb \Bigl(\frac{d\xx^+\!}{d\xx^-\!} \spc \Bigr)^2 - \frac{c}{24\pi } \spc {\rm Sch}(\X^+,\X^-) ,
\ea
which can be thought of as the stress-energy carried by the Unruh radiation that is produced by the moving mirror trajectory specified by $(\xx^+(\tau),\xx^-(\tau))$.  
The reflecting CFT modes and the Unruh radiation produce a force ${\cal F}$ on the boundary, which needs to be taken into account. 
This force ${\cal F}$ must be perpendicular to the trajectory 
\ba
\label{perp}
{\cal F}_{\!+}\spc  \frac{d \X}{d\tau}^{\! +}\!+\spc {\cal F}_{\! -} \spc  \frac{d \X}{d\tau}^{\! -} \is 0 
\ea
to preserve the proper time condition \eqref{proper}. An explicit expression for ${\cal F}$, consistent with the two conditions \eqref{reflect} and \eqref{perp}, was found in \cite{Chung:1993rf} 
\ba
\label{force}
{\cal F}_\pm\is \;  \pm \spc \lb T_{\pm\pm} \rb \spc \frac{d\xx^\pm\!\!}{d\tau} \,  \pm \, {c\over 24\pi} \frac{d^3 \xx^\pm}{d\tau^3}.
\ea
The first term on the r.h.s. represents the force due to the classical reflection of the CFT modes; the second term is a quantum recoil effect due to the  Unruh radiation. The above formula holds for arbitrary boundary trajectories.

We can represent the recoil effect of the Unruh modes as an extra term in the CFT effective action \cite{Chung:1993rf} 
\ba
\label{bott}
{\cal S}^{\rm eff}_{\rm CFT} 
&  \supset &  \frac{c}{48\pi} \!\int \!\! dt \nspc \log\Bigl(\spc \frac{\raisebox{-.5pt}{$d \X^{\!\smpc +}$}\!\!}{\raisebox{.5pt}{$dt$}}\,\spc\Bigr)\!\smpc\frac{d}{dt} \! \log\Bigl( \frac{\raisebox{-.5pt}{$d \X^{\!\smpc -}$}\!\!}{\raisebox{.5pt}{$dt$}}\,\spc\Bigr)^{\strut}_{\strut} . 
 \ea
One easily verifies that the variation of the above effective action produces the extra recoil term in the equation of motion \eqref{perp}-\eqref{force}. Note that both are conformal invariant. The action \eqref{bott} is closely related to the so-called Bott cocycle of the Virasoro group \cite{bott}.

As before, we now introduce a scale $\epsilon$ by assuming that the boundary trajectory is restricted to take the form \eqref{bbv}.
Upon inserting \eqref{bbv}, the effective action \eqref{bott} and its equations of motion \eqref{perp}-\eqref{force} reduce to the Schwarzian action  and equation of motion \eqref{jerk}, with $C = \frac{c \spc \epsilon}{12\pi}$. 

\vspace{-2mm}

	\def\raw{\rightarrow}
\subsection*{Finite temperature}	
\vspace{-3mm}

It is instructive to consider the system at finite temperature  $T = {1}/{\beta}$. In this case, the leading order contribution $\sS_0(u,v)$ to the vacuum entanglement entropy  
across the point $(u,v)$ is given by
\ba
\sS_0(u,v) \is
\frac{c}{6} \log\left(\frac{\frac\beta \pi  \sinh\frac\pi \beta (x^+\! - \! x^-)}
{{2\epsilon} \sqrt{\partial_u x^+ \partial_v x^-}} 
\right) , 
\ea
which via \eqref{sscale}-\eqref{ccmetric} again defines a constant curvature metric. To extract the first order contribution to the vacuum entropy $\sS_1(x)$, we again perform a small coordinate shift 
$x^\pm \to x^\pm\mp \delta \epsilon$ 
which induces a variation $\sS_0(x)\to  \sS_0(x) + \sS_1(x)$ in the leading order vacuum entropy, with 
\ba
\label{stshift}
 \sS_1(x) \is - \frac {4\pi^2 \delta C}\beta  \coth\left(\frac\pi \beta (x^+\! - \! x^-)\right) 
\ea
with $\delta C$ given in \eqref{dc}.
The profile \eqref{stshift} of $S_1(x)$ coincides with the dilaton profile in JT gravity in a black hole background.
We interpret $\sS_1(x)$ as the entanglement entropy between the extra boundary layer with thickness $\delta \epsilon$ and the CFT modes in the region to the left of the point $(x^+,x^-)$. The negative sign means that this entropy is removed in the RG step from $\epsilon \to \epsilon + \delta\epsilon$.

The total thermal entropy of the boundary layer is found by taking the limit $x^+\!-x^- \to \infty$ in equation \eqref{stshift} and integrating the result with respect to the cut-off
\ba
\sS_{b, \rm thermal} \! \is\! \frac{4\pi^2 C}{\beta} = \frac{\pi c \spc \epsilon}{3 \beta} \, . 
\ea
This thermal entropy appears with a minus sign in $S_1(x)$, since it gets subtracted in the RG step.
The middle expression coincides with the thermal entropy of the Schwarzian \cite{Maldacena:2016upp}; the expression on the right equals the thermal entropy of a 2D CFT of central charge $c$ on an interval of width $\epsilon$. 
This result is consistent with the interpretation of Schwarzian QM as the effective 1D theory of a small boundary layer of a 2D CFT at large central charge \cite{Mertens:2017mtv}.

\vspace{-2mm}

\subsection*{\bf Concluding comments}

\vspace{-3mm}

We have shown that, in a 2D CFT with a boundary, the entanglement entropy $S(u,v)$ and modular Hamiltonian $K(u,v)$ associated with a point $(u,v)$ in the bulk, as shown in Figure 1, satisfy local equations of motion that take the same form as those of Jackiw-Teitelboim dilaton gravity coupled to the CFT. In the bulk, the dilaton profile is fixed by the gauge constraints and the 	coupling to JT gravity does not modify the CFT dynamics. 
{Rather, the equations of motion of JT gravity merely dictate the identification of the dilaton with the collective mode $S_1(u,v) = \lb K(u,v)\rb$  of the CFT.} 

We derived the effective equations of motion of the boundary via entropy considerations and energy-momentum conservation, and found that they coincide with those of Schwarzian QM with coupling constant $C = \frac{c\spc \epsilon}{12\pi}$, with $\epsilon$ the distance between the dynamical boundary and the AdS${}_2$ boundary. These equations of motion \eqref{jerk} can be read in two ways: (i) as determining the boundary trajectory in terms of the incoming and outgoing energy-momentum flux, or equivalently, (ii) as determining the outgoing energy-momentum flux for a given boundary trajectory and incoming energy-momentum flux. In the first interpretation, we can view the boundary trajectory $\tau(t)$ as an effective low energy degree of freedom produced by the CFT dynamics. Our result that its dynamics is governed by the Schwarzian action can be used to derive the Lyapunov behavior of the CFT, similar to \cite{Turiaci:2016cvo}.

\vspace{-2mm}

\subsection*{\bf Acknowledgements}

\vspace{-3mm}

We thank Jan de Boer, Bartek Czech, Dan Harlow, Aitor Lewkowycz, Thomas Mertens,  Douglas Stanford, Joaquin Turiaci and Zhenbin Yang for useful discussions and comments. The research of H.V. is supported by NSF grant PHY-1620059. N.C. is supported by  the Research Foundation-Flanders (FWO Vlaanderen).

\end{document}